\documentclass[preprint,aps,a4paper]{revtex4}

\newcommand{\beq}{\begin{equation}}
\newcommand{\eeq}{\end{equation}}
\newcommand{\barr}{\begin{eqnarray}}
\newcommand{\earr}{\end{eqnarray}}

\usepackage{graphicx}
\usepackage{epsfig}
\usepackage{dcolumn}
\usepackage{bm}

\begin{document}

\title{
On the objective existence of physical processes: a physicist's philosophical
view
}
\author{J. Sanchez-Guillen and R.A. V\'azquez}
\affiliation{Departamento de F\'{i}sica de Part\'{i}culas, Campus Sur\\
  15782 Santiago de Compostela, Spain}

\begin{abstract}
Inspired by philosophical ideas of Boltzmann, which are briefly recalled, we
provide strong support for the possibility and convenience of a realistic
world picture, properly nuanced. The arguments have consequences for the
interpretation of quantum mechanics, and for relevant concepts of quantum
field and string theory, like monopoles and branes. Our view is illustrated
with a cybernetic analogy and complemented with a summary of the basic
philosophical concepts.
\end{abstract}

\maketitle

\section{Introduction}

It is common in the Physics literature to find more or less involved
statements on the existence of the basic objects and of their representations.
Relevant examples today may be strings or branes, as atoms and quarks were in
the past.  As a practical or more conscious attitude, the positions range from
two extremes. On one side, existence is granted or denied to the items if they
can be detected (like atoms or quarks) or disproved, like the ether or the
planet Vulcan, without further thoughts about the reality of their
mathematical representative.  In the other extreme \cite{Tegmark} existence is
assigned even to the representations themselves, with varying degrees of
commitment to the philosophical concepts.

In a way this parallels the extremes of the oldest philosophical debate,
between rationalists and empiricists, the most radical forms of which are
naive realism and constructivism. Practitioners of physics are usually closer
to the former view.  Nevertheless one cannot always ignore the philosophical
background, and certainly not concerning the foundations of quantum mechanics,
where the interpretation, which is a basic ingredient, has always been a
matter of debate, often called the problem of realism. This debate can have
also practical implications e.g. for quantum information or quantum
cosmology. In any case, thinking about these abstract matters certainly opens
our vision and improves our questions and understanding.

Our purpose is to provide a clear and simple conceptual (philosophical)
 picture of these important questions, which could also be useful in practice,
 clean of technicalities. Our view was mainly built thinking about the
 philosophical ideas of Boltzmann. They are contained in his only specific
 technical article \cite{Boltz}, which is both long and hard to read, in his
 notes for the Philosophy lectures  held by
Boltzmann in his last years \cite{Fasol}, which are even harder, and
 also informally scattered in his Popular Writings and other popularizations,
 which are very rewarding \cite{Popmodel}.
 
 In the next section \ref{theproblem} we address the main problem of the
 existence of the physical world and its description, beginning with a
 historical perspective to provide a selection of Boltzmann's ideas, and we
 reassure the reader immediately with the basic conclusion, that one can
 safely maintain a realistic position on the objective existence of the
 external world, in permanent evolution, adjusted and regulated by
 experimental confrontation. In Boltzmann's words {\it ``we must adopt the
   objective point of view'', as the phrase goes.}\cite{Boltz}. This will lead
 us to our main proposal, to call existing or non existing only what is
 detectable or falsable respectively. As for the different concepts, at
 various levels of complexity and abstraction, for which neither option can be
 achieved at present, one should distinguish between the physics approach with
 mathematical analysis and measurements, and the philosophical reasoning,
 involving qualitative attributes which can be related to existence.
 Interesting examples of the former are entanglement or virtualities, and of
 the later unavoidability, which is discussed in Section \ref{framework}.
 There we apply these ideas to specific concepts of quantum field theory,
 string theory and cosmology, like monopoles and branes and to the general
 problem of the interpretation of quantum mechanics. Section \ref{conclusions}
 summarizes the conclusions and outlook. In Appendix 1 we illustrate the
 mental representations with a model analogy from neural networks and
 programming languages.We include another Appendix 2 to explain in a concise
 way the philosophical background, relevant for our arguments and beyond.

\section{The general problem of the existence of the external world}
\label{theproblem}

To put the problem in a useful historical perspective, we remind that
Boltzmann, who had based the (mechanical) understanding of thermodynamics on
atoms, had to discuss and defend them frequently against the radical
philosophical positivism as well as against extreme phenomenological
theorists.  The present work is in part an attempt to make more accessible
his ideas and to start applying them to present physics.  The first point of
Boltzmann Philosophy was the necessity to define clearly the concepts
discussed and in fact this is how his purely philosophical work begins and
ends, illustrated with examples and personal anecdotes. This claim for clarity
is a constant in his writings, urging to prevent the perverse antinomies of
philosophy.This will be present in the discussion on the {\it existence} now
and we shall keep it in mind in the applications below, especially to the
interpretation of quantum mechanics.

  In his article Boltzmann goes immediately after this introduction to
describe the process of human perception of the external world, starting from
the observation\footnote{It could be seen as a non trivial extension of {\it
``Cogito ergo sum''...} with {\bf ...deinde sunt}.}  that {\it the laws
according to which our own perceptions run their course are familiar to us and
lie ready in the memory. By attaching these same pictures also to the
perceptual complexes that define the bodies of others, we obtain the simplest
description of these complexes}.  This is elaborated further arguing that in
the extreme idealistic position {\it the sensations and volitions of all
others could not be on the same level as the sensations of the observer, but
would have to be taken as merely expressing the equations between his own
sensations}. The idea is stated more clearly in the next page: {\it Therefore
we designate these alien perceptions with analogous mental signs and words to
those for our own, because this gives a good picture of the course of many
complexes of sensations and this simplifies the world picture}. For clarity
and economy, Boltzmann claims after another couple of pages that {\it we must
adopt the ``objective point of view'', as the phrase goes. It turns out that
the concepts we linked with ``existing'' and ``non existing'' largely remain
applicable unchanged. Those people or inanimate things that I merely imagine
without being forced to do so by the regularities in complexes of perceptions
do not exist for others, they are ``objectively'' non-existing.}

The main conclusion of this line of thought, based largely on the {\it common
 judgement of all}, implies in (our) in simple terms, that one can maintain a
 realistic position, assuming the objective existence of the physical world,
 in a reasonable degree of agreement with our representations thereof, which
 are sufficiently universal and which may evolve as required by the
 experimental confrontation and of course by our own evolution, an essential
 ingredient of Boltzmann Philosophy. For instance he wrote:
{\it The brain we view as the apparatus or organ for producing
 world views, an organ which because of the great utility of these views
 for the preservation of the species has, conformably, with Darwin's theory,
developed in a man to a degree of particular perfection just as the neck in
 the giraffe and the bill in the stork have developed to an unusual length} 
\cite{Boltz}.  

Let us explain briefly the argument, which being {\it philosophical}, has to
use logic starting, of course, from our mental representations, the
concepts. We recognize them and decide whether they are relevant or not from
confrontation with the representations of others. But those are also external
{\it objects}, so that any statement and mental construction is actually based
on the external world, as represented with enough degree of fidelity and
universality by our concepts. The adequacy or correspondence of the reasonably
universal representations (concepts) is based on confrontation and guaranteed
by evolution, which also renders the process dynamical (over large time
scales). Therefore it would be {\it inconsistent} to {\it deny } the reality
of the external world. Notice that one has not {\it proved}  its existence,
but the absurdity of the attempts to disprove it, thereby establishing the
possibility and convenience of an objective world picture.

Boltzmann uses then an ingenious argument to maintain this universal realism
for any kind of brain process and beyond, applying these reasoning
successively to simpler and simpler organisms, reaching the virus and
molecular levels, until confrontation or detection is generalized to
interactions.  Our starting point, the mental processes, beginning with the
primary physical inputs, which are later elaborated in different degrees of
robustness and complexity, yielding the representations with varying degrees
of fidelity and universality, should become ultimately also a question of
bio-physical interactions.  But for our philosophical arguments suffice to say
in that respect that the agreement between nature and reason is because reason
is natural and not the opposite. In Appendix1 we provide a cybernetic analogy
of the cognitive process, which can be useful.

In any case, from the arguments so far, i.e. from clear conceptions, rigour and
logic, one has established the possibility of an "objective" world view, and
the 
convenience of this representation, provided it is sufficiently contrasted and
updated. This point of view can be seen a golden middle between the two
extreme positions, as illustrated in Appendix2.

This leads further to propose, that we call {\it existent} only those
representations which, clearly defined, are physically realizable and
detectable (in principle with some energy transfer \cite{Tait}). A
consequence is that there is no place for gradations in this clear, but
restricted notion of existence: representations which fulfil it correspond to
existing objects, like atoms or neutron stars, while those which do not,
should not be called existing. This notion of existence has many advantages,
like a highest degree of universality. {\it The assumption of different
  degrees of existence would be decidedly inappropriate} as Boltzmann says,
and {\it the denotation must always be so chosen that we can operate
  with the same concepts in the same way under all circumstances, just as a
  mathematician defines negative or fractional exponents in such a way that he
  can operate with them as with integral ones} \footnote{This example
  illustrates the so called accumulative linearity of scientific progress,
  not so clear in Philosophy.}. This avoids confusions, like most of the
dreadful antinomies of philosophy, but it also rises the following problem.

 There are concepts which can be clearly disproved, like ether or the planet
Vulcan. But for many ideas one is not able, at least for different time
periods, to detect or disprove them as defined.  What can then be said about
such useful representations, with respect to the external world?  We have to
address the problem, because as we said, any statement of any kind involves
the representations.

This marks a line between a physics approach, where one has to look for a
verification or falsification, and the philosophical, where one can envisage
attributes,which can be related to existence more or less directly on the one
hand, but which have the possibility of qualitative, more or less coarse
gradations, on the other.  They can be vague, like clarity, simplicity or
beauty, or very sharp, like {\it (in)dispensability}, which is discussed in
the next section at length because of his potential relevance.

A first consequence of this is the great convenience of distinguishing in
physics between qualitative concepts, but which are ultimately philosophical
interpretations, from genuine physical proposals, always falsable in a way or
another. Textbook examples are many of the different proposals to render
quantum mechanics {\it complete} or more understandable, which will be
discussed in \ref{qm}. There is of course place in Physics for useful
qualitative discussions, even in our restrictive philosophical view, as we
discuss next. 

\section{A philosophical Framework for the existence in Physics }
\label{framework}

\subsection{The unavoidable attributes related to existence}
One of the simplest and useful examples of such general predicates related to
 existence can be {\it indispensability} ( or the closely related concept of
 {\it unavoidability}), introduced by Boltzmann less formally for the concept
 of the atom in his popular writings. They are not present in his technical
 technical publication, and so their analysis below, is (needless to say!),
 essentially ours.

Boltzmann argued in his talks and popular writings, that his ideal atoms were,
not only useful, but indispensable. They became of course properly existing
after Einstein computed (following Boltzmann's prescription for fluctuations)
the observed Brownian motion of pollen and made predictions confirmed later by
Perrin \cite{Lindley}. At this level the concept of atoms was defined simply
as elementary grains of matter. Of course concepts have to be defined
precisely, and that of atom was finely sharpened later.  More specifically,
they are represented by complex functions, solutions of linear differential
equation (Schr\"odinger's), which in turn can be combined, enlarging at will
their potential manifestations, as discussed in \ref{qm}.

To illustrate this further let us use the concept of gen, similar in some ways
 to the atom. With a very general definition, as units of transmittable
 information, one can of course call them existing, after their molecular
 structure was found in 1953, but in hindsight one could have shown them to be
 unavoidable, at least since W. Sutton named them in 1922 as the Mendelian
 units of transmission.  Life is even more difficult to define than gen, but
 we think it should not be difficult to show it to be unavoidable under rather
 general conditions. As for conscious life, it seems to us almost hopeless at
 present to define and accordingly much more difficult to argue its
 unavoidability, as Drake ``equation" shows.

Notice the subtle difference between indispensability (Unentbehrlichkeit),
which refers to the object ${\rightarrow}$ concept direction, the ``easy way"
according to Boltzmann, and avoidable (Vermeidlich in German or evitabile in
Latin languages)  more related to the concept ${\rightarrow}$ object direction,
which is harder the more complex the concept, as is clear and it is
illustrated in Appendix2. This admittedly exaggerated subtlety, can
nevertheless illustrate the impossibility of a perfect one to one
correspondence between concepts and the external world, as we shall see is
required by some attempts to make quantum mechanics ``complete''.

 Still in biology let us remind of another example given by Boltzmann for non
existent concepts in the philosophical article: the unicorn. It turns out that
today one can speak of the realizability of such a concept, and in fact it has
been done already, e.g. with drosophila flies. It seems on the other hand
unlikely to be neither unavoidable from evolution, which is difficult anyway,
nor at least stable under it. This can illustrate the role of evolution in the
notion of existence as stated above, and of the relevance of dispensability.

Another interesting example, back to physics, is the concept of the
electromagnetic field, (missed by Boltzmann). Of course the concept of
electric {\it and} magnetic fields, became almost unavoidable after Hertz
discovery, and certainly with the disproof of ether. But after the success of
quantum gauge theory the concept of electromagnetic field, {\it the vector
potential}, is clearly unavoidable. This shows the new level one reaches when
a concept is defined in terms of mathematically deeper theories, as the
fundamental interactions and constituents.

Our next discussion involves in fact concepts, monopoles and branes, which are
expressed mathematically, with higher degrees of abstraction and complexity.
Of course the first question in physics is whether the objects can be
realized, or detected, i.e. registered in processes involving some energy
transfer and which can be reproduced. The properly defined concept will
correspond in that case, and only in that case, to an existing object, or,
shortly, {\it exists} (in the sense of the definition).  Until this can be
achieved one can discuss questions as how fundamental or {\it effective} is
the corresponding mathematical theory, but one can also make progress from a
more philosophical point of view as the analysis of the attribute of
indispensability, applied to the following  relevant examples shows.

The concept of monopoles corresponds to the sources of the magnetic field,
i.e. magnetic charges. They were shown by Dirac to be a way of implementing
discreteness of the electric charge, which requires regularization of
singularities and conservation of symmetries. Its existence has not been
proven so far, although there is a more recent claim in condensed matter
physics experiments, with a special composite called {\it spin ice}
\cite{Castelnovo}. It is a matter of debate at present as to whether these
objects fulfil the requirements of the general class required for discrete
charges. On the other hand there are arguments in favour of the
indispensability of monopoles: fundamental theories with compact (gauge) phase
symmetries, the so called grand unified, imply trivial discreteness of charge,
but at the same time, they also predict monopoles. Grand Unification of
interactions could be established soon. In fact neutrino masses provide a good
hint. This example also shows the importance to consider, as mentioned above,
how fundamental is an object, distinguishing monopoles as extended solutions
of a fundamental theory from those aggregates of particles combined in atoms,
molecules, and further structures.

As the name indicates, the concept of brane \cite{Polchinski} refers to
extended objects in spatial and temporal directions, introduced or appearing
in some string and gravity theories. They are useful for combining gauge
fields and gravity at the quantum level, at least in some approximations, and
in cosmology. They serve so far an auxiliary role. To decide about their
existence one has to consider their energy (or density) and propagation. In
fact they provide a way to implement energy conservation for the strings,
which mediate their interactions, to make particles or even the universe.
Direct observable consequences, to decide if they exist (in a specified class)
are very difficult (for instance, there have been proposals for special
gravitational waves). Alternatively, one could consider if they are
unavoidable from their ability to change the rate of expansion, at present and
in the past (inflation).  But these are not so well understood \cite{Sarkar}. 
So, in contrast to the case of monopoles, we could have to wait
very long to decide about the existence of branes and even about their
indispensability.  Therefore, it would be convenient on occasion to keep this
in mind speaking, or writing about these most interesting concepts.  Another
conclusion of this section is that although indispensability or unavoidability
are at another level (philosophical) than the physical existence, which
requires experimental confirmation or falsification, they can be useful even
in physics. Besides, they are more flexible and admit with full right loose
gradations as {\it almost} or the celebrated {\it for practical purposes}.
From a formal point of view, they could be seen as a much weaker form than the
mathematical attribute {\it necessary}, as is appropriate in Physics where
experiments ultimately decide.

\subsection{Interpretation of Quantum mechanics}
\label{qm}
In quantum mechanics one incorporates the representation from the beginning
which may be one of the reasons for its astonishing performances. One works in
fact at the level of representations without actual reference to the external
world until measurement. These representations are complex "wave" functions,
which can be superposed with the interference properties of waves. Our
proposal requires not to call these representations {\it existing} until they
have been realized in a detectable way, with a probability given of course by
the norm of the combined function. This way one keeps a universal meaning of
the concept and avoids the potential confusion of many interpretations which
have been proposed to cope with the apparent conceptual difficulties of quantum
mechanics. These are mainly the essential probabilistic nature of the
description (``stochastic unit samples''), the mechanism and the nature of the
transition from an uncertain or fluctuating state to the robust and  certain
measurement result ( ``collapse and decoherence'').  Interpretations which in
a more or less subtle way attribute existence to the representations, the wave
functions and their combinations, or to additional auxiliary functions, are
explained in many excellent textbooks \cite{adler}. Well known examples are the
many worlds of Everett, Hartle's consistent histories and the pilot waves of
Bohm.  It is important to remember there is no way to detect or falsify those
interpretations, so we are in fact at the philosophical level, where the
relevant question should be whether these auxiliary objects, worlds or paths,
are unavoidable. The answer is clearly they are not, rather the opposite,
a view shared by many of the
active researchers using those fundamental aspects of quantum mechanics, like
A. Zeilinger \cite{z}. In fact, it is natural to accept limitations to common
sense imposed by physics, as it happened before with simultaneity or with
indistinguishability.

This illustrates the usefulness of our simple proposal, but we do not claim
 that  
 the situation is completely satisfactory. The question
whether quantum mechanics is a complete theory is alive since Einstein,
Podolsky and Rosen posed it \cite{EPR}. One has to define completeness, and of
course they did: {\it every element of physical reality should have a
counterpart in the physical theory}. This strong requirement is very difficult
to meet, as we have seen from the discussion on the cognitive process.  In
fact, trying to complete the theory to meet the above difficulties one has to cope
with very restrictive theorems about the implications of adding new "hidden"
variables \cite {Bell}, under very reasonable general conditions, which have
been always experimentally confirmed in favour of pure quantum mechanics. There
are even stronger theorems limiting the possibility of such extension from
internal consistency \cite{koch}. 

There are partial but promising solutions to the last of the basic mentioned
problems, decoherence \cite{Deco}, but it is clearly beyond the scope and
space limits of this article to enter into details of these well defined
physics. Let us remind instead that there are in fact other reasons for
insatisfaction beyond these conceptual problems, especially in the
relativistic extensions, where infinities appear in the perturbative
solutions. These divergences are under control, and some even well understood,
which is not the case for the extension to gravitational interactions. Also
there are many basic problems related to strong interactions defying solution
for decades, like the confinement of quarks.This is seen by some as a need to
reformulate the foundations, with new physics \cite{adler, th}, which would be
interesting to study in our framework.  From the simple concept of existence
point of view, there are well defined physics notions to measure virtualities
and quantum entanglements. Let us comment briefly on a special one \cite{Nos}.
It is based on the worldline or proper time formulation of quantum mechanics,
due to Fock, Schwinger, and Feynman. There, one parameterizes quantum
amplitudes with an auxiliary parameter called proper time, which controls the
fluctuations of the quantum state in spacetime, and are represented as a path
integral in that auxiliary space. After some manipulations, the integrals can
be computed by combination of numerical and analytical methods, and the
fluctuations visualized. The virtuality of the process is related to different
random walks, more or less directed, which in turn can be put in
correspondence with a Hausdorff dimension, 2 in the extreme quantum or 1 in
the the classical regime. This is to illustrate that there are concepts, like
virtualities, which can be further analysed with our philosophy proposals and
which could help understanding and even visualizing the physics problem.
 
\section{Conclusions}
\label{conclusions}

Building on Boltzmann philosophical ideas, briefly recalled, we have given
arguments based on general concepts and logic, strongly supporting the
possibility and the convenience of maintaining the objective existence of
physics processes. They imply in turn to {\it restrict} that {\it existence}
to concepts which can have a clear manifestation. For concepts which one is
not able for the moment to submit to such a requirement, the philosophical
analysis can make useful contributions, with attributes related to
existence. We elaborated the simple examples of unavoidability and its
reverse.  They were first illustrated with historical interesting examples of
atoms, fields, and even gens and life and then applied in some detail to
relevant concepts at present like monopoles and branes. In quantum mechanics,
they provide a clarification and criticism of popular
interpretations. More involved analysis have been suggested for future
research, in a fruitful interplay of physics and philosophy. We hope that our
presentation is not perceived as an over-simplification, a possibility we
assumed in the spirit of the practical and direct form Boltzmann declared
indispensable for philosophical argumentations.

\section*{Acknowledgements}

JSG has benefitted from many conversations with D. Flamm,
philosopher, physicist and grandson of Boltzmann and with J. Ordo\~nez, who
provided many of the original references. We were encouraged by the late Julio
Abad in previous stages of this work, which will appear in a special
volume in his memory. This was supported in part by Ministerio de Ciencia e
Innovacion and Conselleria de Educacion under Grants FPA2008-01177 and
2006/51.

\section{Appendix1:Basic picture of thought with cybernetics analogy}
Our arguments started from the basics of the cognitive process, where a
physical input triggers a primary signal, which is later processed into a
memory or commemorative recordings of various degrees of complexity, the
concepts, involving neural networks in specific areas in the brain.  They are
decanted trough permanent confrontation with the external world and those of
others, as explained, and become ideas at various degrees of abstraction and
distance from the primary trigger \cite{Cajal}. This basic conception (and
logic) is what has been used for the argument supporting an objective world
picture, in which those representations, the concepts, can be appropriately
said to correspond to existing objects, provided they are detected, or not, if
experimentally disproved.

The human brain produces general frames abstracting relevant properties of
objects, the languages, consisting of words, syntax, and meaning, of different
levels of abstraction and, correspondingly, of retroprojectivity to the
external objects. The main areas and processes involved are becoming known,
but of course there is a long way to go to master the codes, and it is even an
open question whether the whole process can be controlled in a relevant way.
Besides, we are aware of the key role of chemistry in the brains functions!
But for our philosophical reasoning here we only need the general concept,
which can illustrated with a neural network analogy.

In neural networks, one has a first layer of devices (neurons) receiving an
input. This in turn triggers an output from different numbers of neurons to
the next layers, with different weights. Combining outputs, simple functions
(say tanh) connecting two values (on-off) one can produce complicated
functions, which will implement effective operations, like recognising voices
or pictures. In terms of computers these simple functions could correspond to
{\it machine languages}. But one has also {\it object oriented languages},
like C++, where one works with abstract functions (for instance, templates),
and can operate with them. In our analogy, these correspond to more abstract
and general languages, the most universal of which is mathematics.

\begin{figure}[ht]
\begin{center}
\includegraphics[angle=0, width=8.cm]{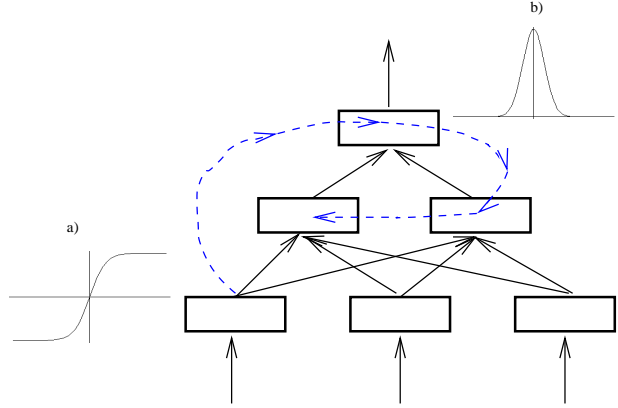}
\caption{Schematic representation of a neural network. The input passing
  through simple functions like (a) is transformed into the output (b). In
  blue (dashed lines) a possible transversal and reverse connection is shown.}
\label{fig1}
\end{center}
\end{figure}

Let us look at the concept of {\it pair}. In the first case, in machine
related language, one has pairs of concrete objects (integers, for example),
which allow in turn for operations like permutations or orderings. In the
second more advanced case, templates provide abstract pairs which can in turn
be paired or otherwise operated successively, including combinations with
other objects. In the neural network this could correspond to nonplanar and
transverse connections, which, require a much larger size and plasticity, like
in parallel vector networks, as indicated schematically in Fig.1

\section{Appendix2: Philosophy background}

As our main conclusion can be seen as a compromise between empiricism and
idealism, we explain these basic philosophical ideas and terminology,
condensed and simplified, using physical examples.

Kant is a standard reference \cite{KO} and for good reason, as mathematics and
physics were his starting point, and it is the first manifestly critical
approach to the theory of knowledge. His position was the first reasonably,
partially realistic one.  In the old debate of empiricism, denying reality of
ideas and the external world and rationalism, assigning it to both of them,
Kant's proposal is a  middle solution: he did not grant general
existence to ideas but he defended some existence of the sensorial world.
Boltzmann's position can be seen as a big step forward in this direction, with 
a sound scientific basis, including evolution and  opening it to future
progress.

Back in 1781 \cite{bi}, Kant noticed that statements can be {\it
analytical} (e.g. the electrons with up or down spin directions in a factorized
tensor product state) or {\it synthetical} (e.g. the electrons in a
symmetrized coupled (entangled) state with a given total spin, $0$ or $1$
). More schematically:``$a$ is in $ab$'' is analytical while  ``$c$
is in $ab$''is synthetical. 

All empirical (called a posteriori) statements are synthetical (experience
always teaches), but the opposite is not true, not all synthetic statements
are empirical: there are some statements with new properties in addition to
the premises (synthetic) which are true independent of empirical experience (a
priori). This simple scheme was thought to apply to mathematics and even some
concepts of physics but it has been generalized (and relativized) by modern
Philosophy of Science.

A proper analysis is also the task of neuroscience and the theory of knowledge
as discussed, but as a philosopher Kant argued that there are preconditions in
humans, such as time and space, which universally allow such processes.  Of
course Kant, a devoted Newtonian who had himself worked out the notion of
galaxies, took time and space as universal and absolute, as well as other
necessary ingredients of thought called {\it categories}.  They are
fundamental ideas like {\it causality}, which had been around from the
beginning of Philosophy and ordered by Aristotle. Needless to say that those
absolute notions were naive, and wrong in strict terms, but the framework was
adequate for scientific discussions and the seed for the later developments.
It is worth reminding ourselves that Boltzmann who proposed a big step
forward, facing the ontological problem \cite{Fasol}, warned against absolute
use of ``laws of thought'' like causality, {\it which we may denote either a
  precondition of all experience or as itself an experience we have in
  conjunction with every other experience}\cite{Boltz}. He also warned
frequently that in the realm of explanations, models and theories could be
also useful, even if apparently wrong.


\begin{thebibliography}{999}
%
\bibitem{Tegmark} M. Tegmark, The Mathematical Universe, 
Foundations of Physics 38 (2008) 101. ArXiv.0704.0646.
%
\bibitem{Boltz} L. Boltzmann, SB Wien Ber. 106 (1897) Part IIa,
  83-109. English version in Brian McGuiness (ed.), Theoretical Physics and
  Philosophical Problems, D. Reidel, Dordrecht, Boston 1974, p. 57.
%
\bibitem{Fasol} I. Fasol-Boltzmann, (ed.) Principien der Naturfilosofi,
Springer Verlag, Berlin (1990).
%
\bibitem{Popmodel} L. Boltzmann, Popul\"ere Schriften, Leipzig (1905)
  Joh. Ambrosius Barth. Models. Encyclopaedia Britannica. 10th and 11th
  editions (1902). 
%
\bibitem{Tait} It is remarkable that when the universal character of energy
  conservation by Helmholtz was still a matter of intense debate, P.G. Tait
  clearly stressed that ``energy conservation merely asserts its objective
  reality'' in his ``Lectures on some recent advances in Physical Science'',
  p.19, MacMillan, 1876, London.
%
\bibitem{Lindley} D. Lindley, Boltzmann's Atoms, The Free Press, New York
  2001. Albeit a popular book, it gives an excellent short account of this
  debate. 
%
\bibitem{Castelnovo} C. Castelnovo, R. Moessner, and S.L. Sondhi, Nature 451
(2008) 42. 
%
\bibitem{Polchinski} J. Polchinski, String Theory, Cambridge University Press,
1998.
%
\bibitem{Sarkar} S. Sarkar, Gen. Rel. Grav. 40 (2008) 269.
%
\bibitem{adler} S.L. Adler, Quantum Theory as an Emergent Phenomenon,
  Cambridge U.P.(2004).
%
\bibitem{z} A. Zeilinger, Einsteins Spuk. Goldmann, Munich (2007).
%
\bibitem{EPR} A. Einstein, D. Podolsky, and N. Rosen, Phys. Rev. 47 (1935)
  777.
%
\bibitem {Bell} J. S. Bell, Phys.Rep. 137 (1986) 49.
%
\bibitem {koch} J. Conway and S. Kochen, Foundations of Physics 36 (2006) 1441.
%
\bibitem{Deco} J.A. Wheeler and W. Zurek, Quantum Theory and Measurement,
Princeton University Press, 1983. 
%
\bibitem {th} G. 't Hooft, ArXiv:hep-th/0707.4568
%
\bibitem{Nos} H. Gies, J. Sanchez-Guillen, and R.A. Vazquez, JHEP 0508 (2005)
067.
%
\bibitem{Cajal} These words are almost literal from
S. Ram\'on y Cajal in ``Textura del sistema nervioso del hombre y los
vertebrados'' (re-edited in 1992 by the Universidad de Alicante, Spain). We
have found \cite{l} that Boltzmann attended his lectures about it in
Worcester, Mass. in 1899.
%
\bibitem{l} J. Sanchez-Guillen, L.E. Boltzmann, PUZ, Zaragoza (2009). 
%
\bibitem{KO} R. Omnes, Understanding Quantum Mechanics, Princeton
University Press, 1999.
%
\bibitem{bi} I. Kant, Kritik der reinen Vernunft (1781).
%
\end{thebibliography}
\end{document}